\documentclass{emulateapj} 

\newcommand{\bl}[1]{\mbox{\boldmath$ #1 $}}

\slugcomment{Accepted by The Astrophysical Journal}
\shorttitle{Lifetime of the embedded phase}
\shortauthors{Vorobyov} 

\begin{document}

\title{Lifetime of the embedded phase of low-mass star formation and the envelope depletion rates}
\author{Eduard I. Vorobyov\altaffilmark{1,}\altaffilmark{2}}
\altaffiltext{1}{Institute for Computational Astrophysics, Saint Mary's University,
Halifax, B3H 3C3, Canada; vorobyov@ap.smu.ca.} 
\altaffiltext{2}{Institute of Physics, South Federal University, Stachki 194, Rostov-on-Don, 
344090, Russia.} 


\begin{abstract}
Motivated by a considerable scatter in the observationally inferred lifetimes of the embedded 
phase of star formation, we study the duration of the Class~0 and Class~I phases
in upper-mass brown dwarfs and low-mass
stars using numerical hydrodynamics simulations of the gravitational collapse of
a large sample of cloud cores. We resolve the formation of a star/disk/envelope system and 
extend our numerical simulations to the late accretion phase when the envelope is 
nearly totally depleted of matter.
We adopted a classification scheme of Andr\'e et al.
and calculate the lifetimes of the Class~0 and Class~I phases ($\tau_{\rm C0}$ and 
$\tau_{\rm CI}$, respectively) based on the mass remaining in the
envelope. When cloud cores with various rotation rates, masses, and sizes 
(but identical otherwise) are considered, 
our modeling reveals a sub-linear correlation between the Class~0 lifetimes and
stellar masses in the Class~0 phase with the least-squares fit exponent $m=0.8\pm0.05$.
The corresponding correlation between the Class~I lifetimes and stellar masses in 
the Class~I is super-linear with $m=1.2\pm0.05$.
If a wider sample of cloud cores is considered, which includes possible variations in the
initial gas temperature, cloud core truncation radii, density enhancement amplitudes, initial gas density
and angular velocity profiles, and magnetic fields,
then the corresponding exponents may decrease by as much as 0.3. 
 The duration of the Class~I phase is found to be longer than that 
of the Class~0 phase in most models, with a mean ratio 
$\tau_{\rm CI}/\tau_{\rm C0}\approx$~1.5--2. A notable exception are YSOs that form 
from cloud cores with large initial density enhancements, in which case $\tau_{\rm C0}$ 
may be greater than  $\tau_{\rm CI}$. Moreover, the upper-mass ($\ga 1.0~M_\odot$)
cloud cores with frozen-in magnetic 
fields and high cloud core rotation rates may have the $\tau_{\rm CI}/\tau_{\rm C0}$ 
ratios as large as 3.0--4.0.
We calculate the rate of mass accretion from the envelope onto the star/disk system
and provide an approximation formula that can be used in semi-analytic models of cloud core
collapse. 
\end{abstract}

\keywords{circumstellar matter --- planetary systems: protoplanetary disks --- hydrodynamics --- ISM:
clouds ---  stars: formation} 

\section{Introduction}

Constraining the lifetimes associated with different phases of low-mass star formation
has been traditionally one of the goals of research.
In spite of much effort in this field, yet there is a considerable scatter 
among observationally estimated lifetimes of the embedded phase.
Most previous observational studies suggested the duration of the Class~0 and Class~I 
phases to be within 
the  0.1--0.7~Myr range \citep{Wilking89,Greene94,Kenyon95,Visser02,Hatchell07}, 
with similar relative lifetimes \citep{Visser02,Hatchell07},
while some argue in favour of a
significantly shorter lifetime for the Class 0 phase, $\tau_{\rm C0}$=0.01--0.06~Myr 
\citep{Andre94}. In the most recent work involving a large set of YSOs
from different star forming clouds, \citet{Enoch09} and \citet{Evans09} advocated for 
the lifetimes of 0.1--0.2~Myr for the Class~0 sources
and 0.44--0.54~Myr for the Class~I ones.

The aforementioned lifetimes
may vary in part due to different estimation techniques and employed wavelengths 
and in part due to observations being confined to different molecular clouds.
The cornerstone assumption of steady star formation, adopted in most observational studies,
may break down if star formation relies heavily on external triggering, 
as may be the case in $\rho$~Ophiuchi \citep{Visser02}. Furthermore,
the environment and local initial conditions determine how long 
the Class~0 and Class~I phases last.

It is therefore interesting to compare the observationally inferred lifetimes
with those predicted from numerical modeling of the cloud core collapse.
For instance, \citet{Masunaga00} performed radiation hydrodynamics simulations of
the gravitational collapse of spherically symmetric cloud cores and found, based on
the synthetic spectral energy distribution, 
that the Class~0 phase lasts around $2\times10^4$~yr. However, their modeling was 
limited to only two cloud cores.
Some work in this direction was also done by \citet{Froebrich06}, who employed smooth
particle hydrodynamics simulations of the fragmentation and collapse of
turbulent, self-gravitating clouds and found Class 0 lifetimes to be of the order of
$(2-6) \times 10^{4}$~yr, significantly lower than those inferred from most observations.
However, they did not provide estimates for the Class I lifetime, probably due to 
an enormous computational cost, and their lifetimes were inferred from 
the mass infall rates at $\sim$~300~AU rather than from more customary diagnostics
such as envelope masses or spectral properties.

In this paper, we perform a comprehensive numerical study of the duration of the embedded
phase of star formation in upper-mass brown dwarfs and low-mass stars. 
Using numerical hydrodynamics simulations,
we compute the gravitational collapse of a large sample of gravitationally unstable 
cloud cores with various initial masses, rotation rates, gas temperatures, density enhancement
amplitudes, truncation radii, and strengths of frozen-in magnetic fields. 
The paper is organized as follows. In Sections~\ref{model} and \ref{init}, 
we briefly review the numerical model and initial conditions.
In Section~\ref{schemes}, we discuss the adopted classification scheme of YSOs. In 
Section~\ref{lifetimes}, 
we search for possible correlations between the Class~0 and Class~I lifetimes, from one hand, 
and stellar and cloud core masses, from the other hand. 
The effect of varying initial conditions is considered in Section~\ref{initcond}.
The model envelope depletion rates are tested against several empirical functions 
in Section~\ref{depletion}
and the main results are summarized in Section~\ref{summary}.

\section{Model description}
\label{model}
Our numerical model is similar to that used recently to study the secular 
evolution of circumstellar disks and the mass accretion rates 
in T Tauri stars and brown dwarfs  \citep{VB09a,VB09b,Vor09}. 
For the reader's convenience, we briefly summarize the basic concept and equations. 

We make use of the thin-disk approximation 
to compute the long-term ($\sim 2$~Myr) evolution of rotating, 
gravitationally unstable cloud cores. 
We start our numerical integration in the pre-stellar phase, which is 
characterized by a collapsing {\it starless} cloud core, continue into the 
embedded phase, which sees the formation of a star/disk/envelope
system, and terminate our simulations in the late accretion phase,
when most of the cloud core has accreted onto the star/disk system. 
Once the disk has self-consistently formed, it occupies the innermost 
regions of our numerical grid, while the infalling cloud core (the envelope) 
occupies the rest of the grid. This ensures that the mass accretion rate onto 
the disk $\dot{M}_{\rm env}$ is self-consistently determined by the dynamics of 
the gas in the envelope, rather than being simply introduced as a free parameter. 

We introduce a ``sink cell'' at $r<5$~AU and impose a free inflow inner boundary condition.
As the gravitational collapse of a cloud core ensues, the matter begins to freely 
flows through the sink cell. We monitor the gas surface density in the sink cell and 
when its value exceeds a critical value $\Sigma_{\rm cr}$ for the transition from 
isothermal to adiabatic evolution 
(see Equation~[\ref{barotropic}]), we introduce a central star.
In the subsequent evolution, ninety per cent of the gas that crosses the inner 
boundary is assumed to land onto the central star plus the inner axisymmetric disk at $r<5$~AU. 
 The mass ratio of the inner axisymmetric disk relative to the star is a few per cent,
so that most of the accreted mass is accumulated by the star.
The inner disk is dynamically inactive; it contributes only to the total gravitational 
potential and its use is necessary to ensure a smooth 
behaviour of the gravity force down to the stellar surface. The other 10\% of the 
accreted gas is assumed to be carried away with protostellar jets. The latter 
are triggered only after the formation of the central star.

The basic equations of mass and momentum transport in the thin-disk approximation are
\begin{eqnarray}
\label{cont}
 \frac{{\partial \Sigma }}{{\partial t}} & = & - \nabla_p  \cdot \left( \Sigma \bl{v}_p 
\right), \\ 
\label{mom}
 \Sigma \frac{d \bl{v}_p }{d t}  & = &  - \nabla_p {\cal P}  + \Sigma \, \bl{g}_p 
 -{Z\over 4\pi} \nabla_p B_z^2   + {B_z \bl{B}_p \over 2\pi} +  (\nabla \cdot \mathbf{\Pi})_p \, ,
\end{eqnarray}
where $\Sigma$ is the mass surface density, ${\cal P}=\int^{Z}_{-Z} P dz$ is the vertically integrated
form of the gas pressure $P$, $Z$ is the radially and azimuthally varying vertical scale height,
$\bl{v}_p=v_r \hat{\bl r}+ v_\phi \hat{\bl \phi}$ is the velocity in the
disk plane, $\bl{g}_p=g_r \hat{\bl r} +g_\phi \hat{\bl \phi}$ is the gravitational acceleration 
in the disk plane, and $\nabla_p=\hat{\bl r} \partial / \partial r + \hat{\bl \phi} r^{-1} 
\partial / \partial \phi $ is the gradient along the planar coordinates of the disk. 
The gravitational acceleration $\bl{g}_p$ includes the gravity of a central point object 
(when formed), the gravity of the inner inactive disk ($r<5$~AU),
and the self-gravity of a circumstellar disk and envelope. The latter component is found 
by solving the Poisson integral using the convolution theorem. The disk is pierced with a 
magnetic field, which has only the vertical component $B_z$ in the disk and both the vertical 
and planar ($\bl{B}_p = B_r \hat{\bl r} + B_\phi \hat{\bl \phi}$) components at the top and bottom 
surfaces of the disk. Because of our assumption of flux freezing 
and a spatially uniform flux-to-mass ratio $\alpha_{\rm m}$, the vertical and planar 
magnetic field components are easily determined from the relations 
$B_{z}=\alpha_{\rm m} 2 \pi G^{1/2} \Sigma$ and 
$\bl{B}_p= -\alpha_{\rm m} \bl{g}_p/G^{1/2}$, respectively \citep[see][for details]{VB06}.

The viscous stress tensor $\mathbf{\Pi}$ is expressed as
\begin{equation}
\mathbf{\Pi}=2 \Sigma\, \nu \left( \nabla v - {1 \over 3} (\nabla \cdot v) \mathbf{e} \right),
\end{equation}
where $\nabla v$ is a symmetrized velocity gradient tensor, $\mathbf{e}$ is the unit tensor, and
$\nu$ is the kinematic viscosity. 
The components of $(\nabla \cdot \mathbf{\Pi})_p$ in polar coordinates ($r,\phi$) 
can be found in \citet{VB09a}. We make no specific assumptions
about the source of viscosity and parameterize the magnitude of kinematic viscosity using a 
usual form of the $\alpha$-prescription 
\begin{equation}
\nu=\alpha \, c_{\rm s,eff} \, Z, 
\end{equation}
where $c_{\rm s,eff}^2=\gamma {\cal P}/\Sigma$ is the square of the effective sound speed and the ratio
of specific heats is set to $\gamma$=1.4. We note that $c_{\rm s,eff}$ in this formula is 
a dynamically varying quantity and is calculated at every time step during the evolution.

In this paper, we use a spatially and temporally uniform $\alpha=0.005$. This choice is based on 
the recent work of \citet{VB09a}, who studied numerically the secular evolution
of viscous and self-gravitating disks. They found that {\it if} circumstellar disks
around solar-mass protostars could generate and sustain turbulence then the temporally and 
spatially averaged $\alpha$ should lie in the $10^{-3}-10^{-2}$ limits. 
Smaller values of $\alpha$ ($\la 10^{-4}$) have little effect on the resultant mass accretion history,
while larger values ($\alpha \ga 10^{-1}$) destroy circumstellar disks during less than 1.0~Myr of 
evolution and are thus inconsistent with mean disk lifetimes of the order of 2--3~Myr.

Equations~(\ref{cont}) and (\ref{mom}) are closed with a barotropic equation
that makes a smooth transition from isothermal to adiabatic evolution at $\Sigma = \Sigma_{\rm cr}$:
\begin{equation}
{\cal P}=c_{\rm s}^2 \Sigma +c_{\rm s}^2 \Sigma_{\rm cr} \left( \Sigma \over \Sigma_{\rm cr} \right)^{\gamma},
\label{barotropic}
\end{equation}
We note that in this equation (and further in the text) $c_{\rm s}^2=R T_0/\mu$ 
is the sound speed at the beginning of numerical simulations,
which depends on the adopted initial gas temperature $T_0$ of a cloud core and 
the mean molecular weight $\mu$. The value of $\Sigma_{\rm cr}$ is calculated during the numerical
simulations as $\Sigma_{\rm cr}=m_{\rm H} \mu \, n_{\rm cr}\,2 \,Z$, where the critical 
volume number density $n_{\rm cr}$ is set to $10^{11}$~cm$^{-3}$ \citep{Larson} 
and the mean molecular weight is set to 2.33. 
The scale height $Z$ is calculated using the assumption of vertical
hydrostatic equilibrium \citep[see][]{Vor09}. We note that $Z$ is an increasing function of radius,
which makes $\Sigma_{\rm cr}$ to increase with radius as well. In practice, this means
that the inner disk regions are significantly warmer than the outer regions, since the optically 
thick regime in the inner regions is achieved at lower $\Sigma$. 
This in turn impedes the development of gravitational instability and fragmentation 
in the inner disk, in agreement with more 
sophisticated numerical simulations and theoretical predictions
that directly solve for the energy balance equation \citep[see e.g.][]{Boley09}. 

Equations~(\ref{cont}) and (\ref{mom}) are solved in polar 
coordinates $(r, \phi)$ on a numerical grid with
$256 \times 256$ grid zones. 
We use the method of finite differences with a time-explicit solution procedure similar 
in methodology to the ZEUS code, with the advection and 
the force terms treated separately using the operator-split method. Advection is
performed using the second-order van Leer scheme. A small amount of artificial viscosity 
that spreads shocks over
two grid zones is added. However, the resulted artificial viscosity torques are 
negligible in comparison to gravitational and $\alpha$-viscosity ones. 
The radial points are logarithmically spaced.
The innermost grid point is located at $r_{\rm sc}=5$~AU, and the size of the 
first adjacent cell varies in the 0.12--0.17~AU range depending on the cloud core size.

\section{Initial conditions}
\label{init}
We start our numerical simulations from {\it starless}, gravitationally bound cloud cores, 
which have surface densities 
$\Sigma$ and angular velocities $\Omega$ typical for a collapsing, axisymmetric, magnetically
supercritical core \citep{Basu}
\begin{equation}
\Sigma={r_0 \Sigma_0 \over \sqrt{r^2+r_0^2}}\:,
\label{dens}
\end{equation}
\begin{equation}
\label{omega}
\Omega=2\Omega_0 \left( {r_0\over r}\right)^2 \left[\sqrt{1+\left({r\over r_0}\right)^2
} -1\right],
\end{equation}
where $\Omega_0$ is the central angular velocity, 
$r_0$ is the radius of central near-constant-density plateau defined 
as $r_0 = \sqrt{A} c_{\rm s}^2 /(\pi G\Sigma_0)$. We note that when $r\gg r_0$,
the resultant gas surface density becomes $\Sigma=\sqrt{A} c_{\rm s}^2/(\pi G r)$. 
This profile can be derived from the following gas volume density distribution
$\rho =A c_{\rm s}^2/(2 \pi G r^2)$, if the latter is
integrated in the vertical direction assuming a local vertical hydrostatic equilibrium, 
i.e. $\rho = \Sigma/(2Z)$ and $Z=c_{\rm s}^2/(\pi G \Sigma)$. Therefore, our initial density
configuration is characterized by a factor of $A$ positive density enhancement as compared to
that of the singular isothermal sphere $\rho_{\rm SIS} =c_{\rm s}^2/(2\pi G r^2)$ \citep{Shu77}.

Cloud core are characterized by the ratio of
rotational to gravitational energy $\beta=E_{\rm rot}/|E_{\rm grav}|$, where the
rotational and gravitational energies are calculated as
\begin{equation}
E_{\rm rot}= 2 \pi \int \limits_{r_{\rm sc}}^{r_{\rm
out}} r a_{\rm c} \Sigma \, r \, dr, \,\,\,\,\,\
E_{\rm grav}= - 2\pi \int \limits_{r_{\rm sc}}^{\rm r_{\rm out}} r
g_r \Sigma \, r \, dr.
\label{rotgraven}
\end{equation}
Here, $a_{\rm c} = \Omega^2 r$ is the centrifugal acceleration, and $r_{\rm out}$
is the outer cloud core radius. Cloud cores are initially isothermal, with the gas temperature 
varying between $T_0=10$~K and 18~K, depending on the model.

\begin{table*}
\center
\caption{Model parameters}
\label{table1}
\begin{tabular}{cccccccccc}
\hline\hline
Set & $\beta$ & $\Omega_0$ & $r_0$ & $M_{\rm cl}$ & $r_{\rm out}/r_0$ & $T_0$ & $A$ & $\alpha_{m}$ & N  \\
\hline
 1  & $2.17\times 10^{-3}$ & $0.57-3.30$ & $583-3360$  & $0.34-1.9$ & 6 & 10 & 2 & 0 & 8   \\
 2  & $3.20\times 10^{-3}$ & $0.57-5.60$ & $411-4110$  & $0.24-2.4$ & 6 & 10 & 2 & 0 & 12   \\
 2T & $3.20\times 10^{-3}$ & $1.30-10.5$ & $411-3360$  & $0.43-3.5$ & 6 & 18 & 2 & 0 & 11   \\
 2A & $3.20\times 10^{-3}$ & $0.98-9.6$  & $342-3360$  & $0.40-3.9$ & 6 & 10 & 8 & 0 & 12   \\
 2MF & $3.20\times 10^{-3}$ & $0.57-5.7$ & $411-4110$  & $0.24-2.4$ & 6 & 10 & 2 & 0.3 & 12 \\
 2E & $3.20\times 10^{-3}$ & $1.0- 5.7$  & $411-2230$  & $0.5-2.8$ & 12 & 10 & 2 & 0 &  9 \\
 3  & $6.90\times 10^{-3}$ & $1.11-16.6$ & $206-3086$  & $0.12-1.8$ & 6 & 10 & 2 & 0 & 12   \\
 4  & $9.90\times 10^{-3}$ & $1.10-25.0$ & $164-3771$  & $0.095-2.2$& 6 & 10 & 2 & 0 & 10  \\
 5  & $1.34\times 10^{-2}$ & $1.33-35.0$ & $137-3600$  & $0.079-2.1$& 6 & 10 & 2 & 0 & 10 \\
 6  & $2.22\times 10^{-2}$ & $1.56-45.0$ & $102-3940$  & $0.06-2.3$ & 6 & 10 & 2 & 0 & 10 \\
 \hline
\end{tabular} 
\tablecomments{All distances are in AU, angular
velocities in km~s$^{-1}$~pc$^{-1}$, masses in $M_\odot$, initial gas temperatures $T_0$ in Kelvin and $N$ is the number of models in each set.}
\end{table*}

We present results from 10 sets of models, the parameters of which are summarized
in Table~\ref{table1}. Every model set has a distinct 
ratio $\beta$, the adopted values of which lie within
the limits inferred by \citet{Caselli} for dense molecular cloud cores, $\beta=(10^{-4} - 0.07)$.
We note that by construction $\beta=0.91\Omega_0^2 r_0^2/c_{\rm s}^2$ \citep{Vor09b}.
In addition, each set of models is characterized by a distinct ratio $r_{\rm out}/r_0$ 
in order to generate gravitationally unstable truncated cores of similar form. 
For most model sets, we choose $r_{\rm out}/r_0=6.0$.

Every model set contains several individual models\footnote{The exact number of 
these models is 
specified in the last column of Table~\ref{table1}} that are characterized by distinct
masses $M_{\rm cl}$, outer radii $r_{\rm out}$, and central angular velocities $\Omega_0$,
but have equal $\beta$ and $r_{\rm out}/r_0$.
The actual procedure for generating a specific cloud core with a given value of $\beta$ 
is as follows. First, we choose
the outer cloud core radius $r_{\rm out}$ and find $r_0$ from the condition $r_{\rm out}/r_0=const$.
Then, we find the central surface density $\Sigma_0$ from the relation 
$r_0=\sqrt{A}c_{\rm s}^2/(\pi G \Sigma_0)$ and determine the resulting cloud core mass
$M_{\rm cl}$ from Equation~(\ref{dens}). Finally, the central angular velocity $\Omega_0$
is found from the condition $\beta=0.91\Omega_0^2 r_0^2/c_{\rm s}^2$. 

\begin{figure}
  \resizebox{\hsize}{!}{\includegraphics{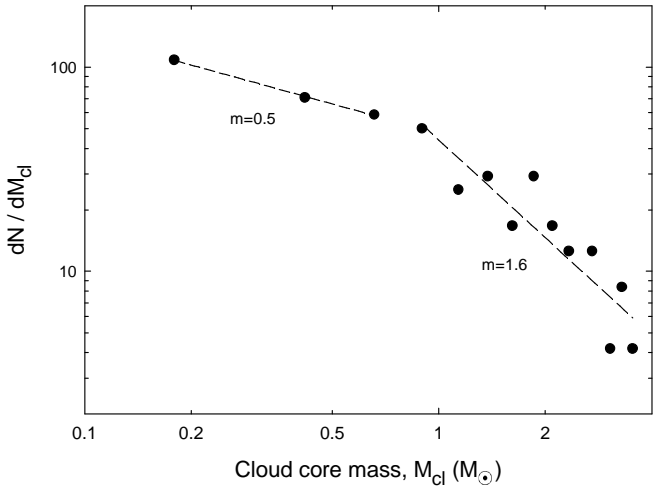}}
      \caption{Initial cloud core mass function $dN/dM_{\rm cl}=M_{\rm cl}^{-m}$ 
      for 106 model cloud cores organized in 14 bins of equal size $dM_{\rm cl}$. 
      The parameters of the cloud cores are listed in Table~\ref{table1}. The dashed lines present
      the least-squares best fits to low-mass cloud cores ($0.06~M_\odot < M_{\rm cl} \le 0.6~M_\odot$,
      left line) and intermediate- and upper-mass cloud cores 
       ($0.6~M_\odot < M_{\rm cl} < 3.9~M_\odot$, right line). }
         \label{fig1}
\end{figure}

In total, we have simulated numerically the time evolution of 106 cloud cores spanning
a range of initial masses between $0.06~M_\odot$ and $3.9~M_\odot$. We have split this mass interval
into 14 bins of equal size $dM_{\rm cl}$.
The resulting initial cloud core mass function $dN/dM_{\rm cl}=M_{\rm cl}^{-m}$ is shown
in Figure~\ref{fig1} by filled circles. The dashed lines present the least-squares best fits to 
low-mass ($0.06~M_\odot < M_{\rm cl} \le 0.6~M_\odot$, left line) cloud cores
and intermediate- and upper-mass ($0.6~M_\odot < M_{\rm cl} < 3.9~M_\odot$, right line) 
cloud cores. The corresponding  exponents are $m=0.5\pm0.1$ and $m=1.6\pm 0.2$, respectively. 
Our model cloud core function is similar to that inferred from nearby star-forming regions
\citep[e.g.][]{Enoch06}. The masses of our model cloud cores have been
selected from a mass distribution that mimics the observed core
mass distribution in order to perform averages of the model parameters (such as disk masses, Class~0/I
lifetimes, etc.) weighted with the cloud core mass function.

\section{Classification schemes}
\label{schemes}
Modern classification schemes of young stellar objects are designed to distinguish between 
the four main physical phases of their evolution: the cloud core collapse and formation of
a protostar and a disk (Class 0 and Class I), envelope clearing and disk accretion onto the
star (Class II), and 
disk dissipation and planet formation (Class III). The spectral energy distribution is often used
to relate a YSOs to a particular class \citep[see][for a thorough review]{Evans09}.
For instance, \citet{Lada87} used the spectral index n (defined by $\nu F_\nu \propto \nu^n$, 
where $F_\nu$ was the infrared flux density at frequency $\nu$) to distinguish between  Class~I, 
Class~II, and Class~III objects. Later, \citet{Andre93} introduced
the ratio of submillimeter to bolometric luminosity $L_{\rm submm}/L_{\rm bol}$ to split the early embedded
stage into the Class 0 and Class I phases based on the mass reservoir remaining in the envelope. 
This separation was originally motivated by the discovery of very young protostars
in collapsing cloud cores that were previously being thought of as starless. 
But it also makes physical sense because the
Class~0 phase can be identified with a period of temporally increasing $L_{\rm bol}$ and Class I
phase with a later period of decreasing $L_{\rm bol}$ \citep{VB05a}.
Finally, \citet{Myers93} suggested the use of a bolometric temperature $T_{\rm bol}$, defined as 
the temperature of a blackbody with the same flux-weighted mean frequency as the actual spectral energy
distribution.

In our case, it is difficult to use any of the aforementioned schemes due to a simplified treatment
of the thermal physics in our numerical simulations. 
Moreover, the large inclination dependence
introduced by the outflow cavities prevents both $T_{\rm bol}$ and $L_{\rm bol}/L_{\rm submm}$ from
being good evolutionary indicators \citep{Dunham10}.
Therefore, we use the classification breakdown suggested also by \citet{Andre93}
and based on the mass remaining in the envelope
\begin{equation}
\begin{array}{ll}
\mathrm{Class~0} \,\,\, & M_{\rm env} \ge 0.5 M_{\rm cl} \\
\mathrm{Class~I} \,\,\, & 0.1 M_{\rm cl} \le M_{\rm env} < 0.5 M_{\rm cl} \\
\mathrm{Class~II} \,\,\, & M_{\rm env} < 0.1 M_{\rm cl}.
\end{array}
\label{Andrescheme}
\end{equation}
According to this scheme (hereafter, AWTB scheme), transition between Class 0 and Class I objects occurs when 
the envelope mass $M_{\rm env}$ decreases to half of the initial cloud core 
mass $M_{\rm cl}$. The Class II phase ensues by the time
when the infalling envelope nearly clears and its total mass drops
below 10\% of the initial cloud core mass. Of course, we acknowledge that these numbers
are somewhat arbitrary and the use of other classification diagnostics may introduce
a systematic shift to our derived lifetimes. 
e spectral energy distribution.

\begin{figure}
  \resizebox{\hsize}{!}{\includegraphics{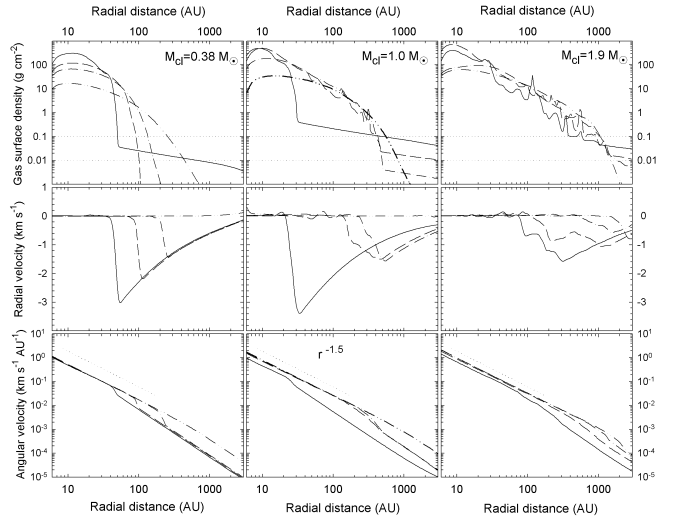}}
      \caption{Azimuthally averaged profiles of the gas surface density (top row), radial
      velocity (middle row), and azimuthal velocity (bottom row) for three models from model 
      set~2.  The initial cloud core masses
      are indicated in the top row. Four different evolution times (since the formation
      of the protostar) are shown with solid (0.1~Myr), dashed (0.2~Myr), dash-dotted (0.3~Myr),
      and dash-dot-dotted (1.0~Myr) lines. The dotted horizontal lines in the top row mark 
      the two adopted values of
      the critical gas surface  density $\Sigma_{\rm
      d2e}$ for the disk to envelope transition. The dotted line in the bottom row presents the 
      angular velocity in the gravitational field of a 4.0~$M_\odot$ star.}
         \label{fig2}
\end{figure}

In order to use the AWTB scheme, we need to know which
part of our numerical grid is occupied by the disk and which part belongs to
the infalling envelope at any time instance of the evolution (note that we know accurately
the stellar mass). This turned out to 
be a nontrivial task. An intuitive step of calculating the rotation profile and separating
between Keplerian and sub-Keplerian regions of the grid sometimes failed to work due to large
radial motions occasionally present in the disk. Besides, the disk is not exactly Keplerian
due to a noticeable contribution from its self-gravity. Therefore, we have developed a hybrid
method that makes use of the radial velocity field in the envelope and a critical surface
density for the disk-to-envelope transition. 

Figure~\ref{fig2} illustrates our idea and shows the 
azimuthally-averaged gas surface density $\overline\Sigma$ (top row), radial 
velocity $\overline{v}_r$ (middle row), and azimuthal velocity $\overline{v}_\phi$ 
(bottom row) distributions as a function of 
radial distance $r$ for three representative models from model set~2. 
The initial cloud core masses $M_{\rm cl}$ are shown in the top row to distinguish between the models.
For each model, the radial profiles at four different evolutionary times 
(elapsed since the protostar formation) are plotted with the solid (0.1~Myr), dashed (0.2 Myr), 
dash-dotted (0.3~Myr), and dash-dot-dotted (1.0~Myr) lines. 
From the radial distribution of $\overline{\Sigma}$ it is seen that the disk is compact and dense
upon formation but later it spreads radially outward due to the action of viscous torques, which 
are positive in the outer disk regions \citep{VB09a}. The disk has a near-Keplerian rotation profile
as can be seen from the comparison of $\overline{v}_\phi$ with the dotted line, the latter 
showing the angular velocity in the gravitational field of a 4.0~$M_\odot$ star.
It is worth noting that the evolution of the unmagnetized models seems to approach a
self-similar behavior in the infalling envelope. This is evident in the $M=1.0~M_\odot$ model,
which radial velocity and surface density distributions approach a free-fall profile proportional 
to $r^{-1/2}$.

In our numerical simulations, the disk occupies the inner regions (small $r$), 
while the envelope---the outer region of the computational grid (large $r$).
A large drop in $\overline{v}_{\rm r}$ (by absolute value) correspond to the radial position where 
the infalling envelope lands onto the disk (the so-called accretion shock).  The shock 
position propagates radially
outward, reflecting the disk expansion. It is also evident that the shock front becomes 
less sharp with increasing cloud
core mass, reflecting an increasingly complex spiral structure of massive disks 
(recall that Figure~\ref{fig2} presents azimuthally averaged profiles).
It is seen that, in general, the envelope is characterized by large negative $\overline{v}_r$ 
due to gravitationally driven collapse and low $\overline{\Sigma}$ due to gas depletion.
On the other hand, the disk has much smaller $\overline{v}_r$ due to 
near-(but~never~complete)-centrifugal balance and larger $\overline{\Sigma}$ due to 
mass accumulation. The outer parts of the disk expand and 
the overall disk density decreases with time  due to ongoing angular momentum
redistribution. In general, we find that {\it massive} disks are not in steady state
but exhibit substantial radial pulsations due to the time-dependent strength of 
gravitational instability in the disk.  
The gas surface density profile declines with radius but some nonmonotonic
behavior is evident for upper-mass models due to fragmentation in the outer parts 
of the disk. 

Having analyzed these data, we adopt the following procedure to
distinguish between the infalling envelope and the disk. 
We march from the outer computational boundary toward
the inner one (in the direction of decreasing $r$) and determine the radial distance 
at which the following two 
criteria are fulfilled simultaneously: $\overline{\Sigma}>\Sigma_{\rm d2e}$ and $\overline{v}_r>0$.
This radial distance is 
set to represent the disk's tentative outer radius $r_{\rm d}$ (at a given time instance of the evolution).
For the critical transitional density $\Sigma_{\rm d2e}$, we choose 0.01~g~cm$^{-2}$ 
or 0.1~g~cm$^{-2}$ shown in Figure~\ref{fig2} with horizontal dotted lines. All material that 
is located within $r\le r_{\rm d}$ is considered to belong to the disk.
This procedure is repeated each time we want to determine the disk and envelope masses and it 
has proven to be most accurate in our circumstances.

\begin{figure}
  \resizebox{\hsize}{!}{\includegraphics{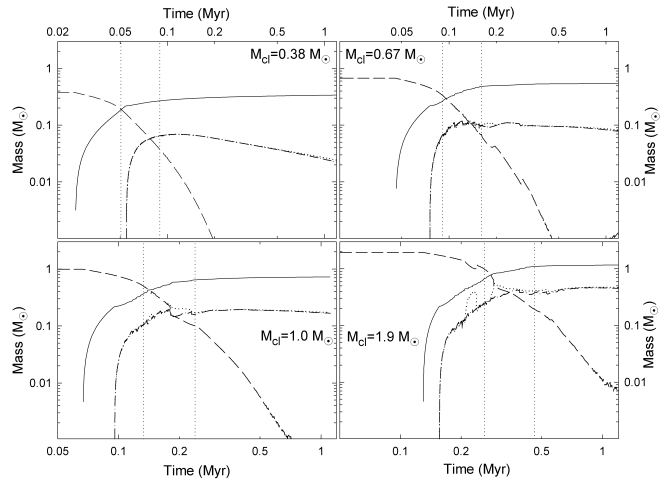}}
      \caption{Time evolution of the stellar mass (solid lines),
      envelope mass (dashed lines) and disk mass (dotted and dash-dotted lines)
      for four models from model set~2. The initial cloud core masses $M_{\rm cl}$
      are indicated in each panel. Two disk masses based on the value of the critical gas
      surface density $\Sigma_{\rm d2e}$ are shown by the dash-dotted lines ($\Sigma_{\rm d2e}=0.1$~g~cm$^{-2}$)
      and dotted lines $\Sigma_{\rm d2e}=0.01$~g~cm$^{-2}$. The vertical dotted lines 
      mark the onset of the Class~I phase (left line) and Class~II phase (right line).}
         \label{fig3}
\end{figure}

Figure~\ref{fig3} presents the integrated disk (dotted and dash-dotted lines), 
envelope (dashed lines), and stellar masses (solid lines) as a function of time 
elapsed since the beginning of numerical simulations 
in four representative models from model set~2. The initial cloud core
masses $M_{\rm cl}$ are indicated in each panel. Two different disk masses are shown based on the 
critical density $\Sigma_{\rm d2e}=0.01$~g~cm$^{-2}$ (dotted lines) and 
$\Sigma_{\rm d2e}=0.1$~g~cm$^{-2}$ (dash-dotted lines).
The vertical dotted lines mark the onset of the Class I phase (left line) and Class II phase 
(right line).
It is seen that the appearance of the star occurs at progressively later evolution times 
with increasing cloud core mass. This is due to the fact that the free-fall time $\tau_{\rm ff}$ 
of a cloud core is linearly proportional to its mass (see the Appendix). Indeed, in the 
$M_{\rm cl}$=0.38~$M_\odot$ model, the star appears at $t=0.026$~Myr, whereas in 
the $M_{\rm cl}$=1.9~$M_{\odot}$ model the star appears 
at $t=0.129$~Myr, which is in good agreement
with what is expected from the linear proportionality between $\tau_{\rm ff}$ and $M_{\rm cl}$.
It is also seen that the difference between disk masses based on the two values of $\Sigma_{\rm d2e}$
is mostly indistinguishable, apart from several episodes when some noticeable mismatch is evident. 
These episodes occur only in the early evolution (and in massive disks) 
and are associated with a transitory disk 
expansion after the accretion of massive fragments (formed due to disk fragmentation) 
by the central protostar \citep{VB06}.

Figure~\ref{fig3} also shows that the disk mass is {\it always} lower than that of the protostar.
This is typical of self-gravitating disks in which self-gravity is computed 
self-consistently \citep{Vor09,Kratter09}, rather than being calculated using an effective 
viscosity proportional to the Toomre $Q$-parameter as in \citet{LP90}. In the latter case, 
the stellar mass may be substantially underestimated in models with large
disk-to-star mass ratios, whereas the disk mass agrees better with that
obtained from self-consistent calculations of disk self-gravity \citep{Vor09c}.
A net result of this discrepancy is that the disk mass starts to considerably 
exceed that of the star soon after its formation, which contradicts observations.
We note that the time difference
between the star and disk formation events is somewhat overestimated in our
numerical simulations due to the use of the sink cell. The rate at which the envelope 
mass declines with time is clearly
dependent on $M_{\rm cl}$---the envelope depletes much slower in models with greater $M_{\rm cl}$.

\section{Lifetimes of the embedded phase}
\label{justify}
In this section, we calculate lifetimes of the Class 0 and Class I phases based
on the AWTB criterion (see Equation~[\ref{Andrescheme}]).
Depending on the adopted value of the critical density $\Sigma_{\rm d2e}$ for the disk-to-envelope 
transition (0.01~g~cm$^{-2}$ or 0.1~g~cm$^{-2}$), the resulting
lifetimes may differ by a factor of unity. In order to minimize the uncertainty, 
we use the arithmetically averaged values. 
For model set~6, however, we use lifetimes based on 
$\Sigma_{\rm d2e}=0.01$~g~cm$^{-2}$. This is because model set~6 produces 
most massive and extended disks (due to considerable rotation rates of the corresponding 
cloud cores), which are best described by the smaller value of $\Sigma_{\rm d2e}$.

For the adopted initial (isothermal) gas surface density distribution of the form 
$\Sigma\propto r^{-1}$, one may expect that Class~0 and Class~I lifetimes ($\tau_{\rm C0}$ and $\tau_{\rm
CI}$, respectively) scale linearly 
with cloud core mass $M_{\rm cl}$. Indeed, for this case, the enclosed mass is proportional to radius
and therefore the crossing-time $t_{\rm cr}=r/c_s$ of an infalling gas shell is roughly 
proportional to the enclosed mass. A more rigorous derivation given in the Appendix shows that
the free-fall time $\tau_{\rm ff}$ is linearly proportional to $M_{\rm cl}$, inverse proportional 
to the initial density enhancement amplitude $A$, and scales with the initial gas temperature as 
$T^{-3/2}$. If the ratios $\tau_{\rm C0}/\tau_{\rm ff}$ and $\tau_{\rm CI}/\tau_{\rm ff}$ 
are independent of cloud core mass (which is not always the case,
see Figure~\ref{fig7}), then a linear relation between
lifetimes and cloud core masses may also be anticipated.

However, the Class~0/I lifetimes are often inferred from observations using stellar characteristics,
and not those of cloud cores. Therefore, it is desirable to relate lifetimes with observationally 
meaningful stellar properties such as stellar masses. The existence of a linear 
dependence between lifetimes and {\it stellar} masses, similar to that anticipated for cloud
core masses, is  
less obvious {\it a priori}. First, the stellar mass varies significantly during the 
Class~0 phase and, to a lesser extent, during the Class~I phase (see Figure~\ref{fig3}). 
It could be observationally difficult 
to know the exact evolutionary instance of the forming star, i.e., whether the star is near the beginning
of, say, the Class~0 phase or in the midway of it. Second, not all infalling mass goes into 
the star. A significant fraction lands onto the disk, and the disk-to-star mass ratio also varies
significantly during the early evolution (Figure~\ref{fig3}, see also \citet{Vor09}). These facts
suggest that a linear correlation between lifetimes  and stellar masses 
can be theoretically anticipated only if the latter are the {\it final} masses. 
On the other hand, using  the final stellar masses in the lifetime--stellar mass dependence may 
be misleading, since observations would provide a whole spectrum of masses at various evolution 
instances of the Class~0 and Class~I phases, and not only at the end of them. 

To complicate things even further, the lifetime--stellar mass correlation may be 
influenced by other cloud core parameters such as magnetic fields and the initial rate of rotation,
the effects of which are difficult to evaluate theoretically. And finally, although the adopted
$\Sigma\propto r^{-1}$ scaling of the initial gas surface density with radius is borne out 
observationally \citep{Bacmann00,Wolf09} and theoretically \citep{Basu}, there is 
no guarantee that other
scaling laws could not be realized in nature. This justifies a detailed numerical 
investigation into to the problem.

\subsection{Lifetimes versus stellar masses}
\label{lifetimes}

\begin{figure*}
 \centering
  \includegraphics[width=18cm]{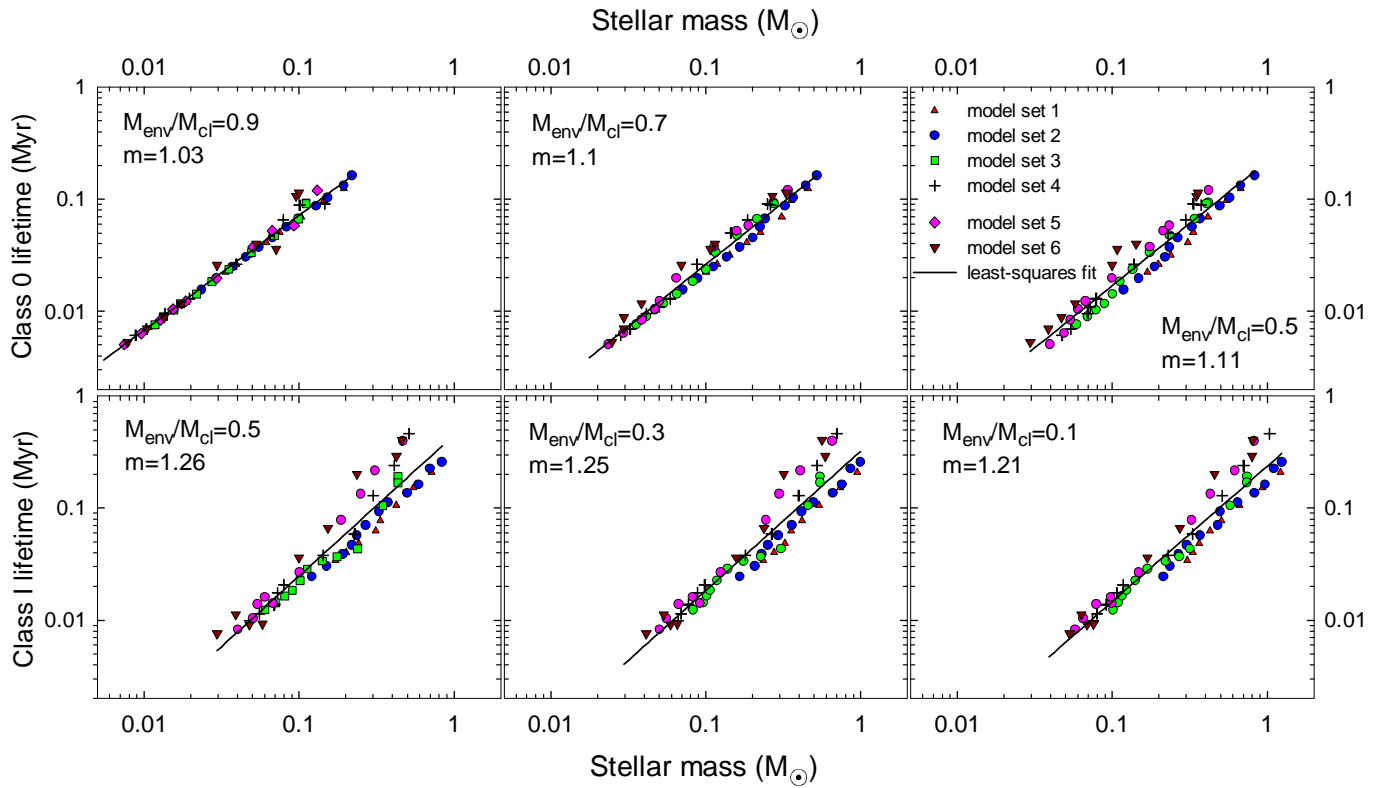}
      \caption{Lifetimes of the Class~0 phase (top row) and Class~I phase (bottom row) versus 
       stellar masses calculated at the beginning (right column), in the midway (middle column), and
       at the end (right column) of each evolution phase. 
       The data for six model sets (see Table~\ref{table1}) are shown. The solid lines present 
       the least-squares fits to these data. The ratio $M_{\rm env}/M_{\rm cl}$ of the 
       envelope mass to the initial cloud core mass and the exponent $m$ of the least-squares fit
       are indicated in every panel.}
         \label{fig4}
\end{figure*}

In this section, we present numerically derived relations between Class~0 and Class~I
lifetimes and corresponding stellar masses, $M_{\rm \ast,C0}$ and $M_{\rm\ast,CI}$. 
Since stellar masses quickly increase with time (see Figure~\ref{fig3}), 
we choose to consider five typical stellar masses based on the ratio $M_{\rm env}/M_{\rm cl}$. 
This ratio is useful because it gradually decreases with time and 
can be used as a tracer of the early evolution of a protostar. 
In particular, we consider three stellar masses derived when the ratio $M_{\rm env}/M_{\rm cl}$ 
becomes equal 0.9, 0.7, and 0.5. These stellar masses are meant to represent typical ones 
in the beginning, in the midway, and at the end of the Class~0 phase, respectively.
For the typical stellar masses in the Class~I phase, we choose three values based on 
$M_{\rm env}/M_{\rm cl}$=0.5, 0.3, and 0.1. We note that we use the same stellar masses at the end of the Class~0 phase and at the beginning of the Class~I phase, but the corresponding 
lifetimes are different. Of course, these five stellar masses 
cannot represent the whole possible spectrum of stellar masses that can be observationally
detected in the embedded phase of star formation. Nevertheless, they can help to reduce
a possible bias towards a particular evolution stage and can help to make our
lifetime--stellar mass relation more observationally meaningful. 

Figure~\ref{fig4} presents our model lifetimes (in Myr) of the Class 0 phase (top panels)
and Class I phase (bottom panels) versus stellar masses (in $M_\odot$) for six model sets. 
More specifically, model set 1 is plotted by red triangles-up, 
model set 2---by blue circles, model set 3---by green squares, model set 4---by plus signs, 
model set 5---by pink diamonds, and model set 6---by brown triangles-down.
Stellar masses in each panel are derived at specific evolutionary times as specified
by the ratio of envelope mass to cloud core mass $M_{\rm env}/M_{\rm cl}$. 
In particular, the upper-left panel employs stellar masses obtained at the beginning of 
the Class~0 phase, respectively. The upper-middle and upper-right 
panels use stellar masses derived in the midway and at the end of the Class~0 phase.
The same age segregation is used in the bottom row of panels only now the stellar 
masses are related to the Class~I phase.

Each symbol of same color and shape within a given set of models represents an individual 
object, which has formed from a cloud core of distinct mass, 
rotation rate, and outer radius.

There are two interesting features in Figure~\ref{fig4} that deserve particular attention.
First, and most important one, is a notable dependence of the derived lifetimes on 
the mass of the central object. The least-squares best fits (solid lines) yield a near-linear
correlations between the Class~0 lifetimes and stellar masses and a somewhat steeper
correlation for the case of the Class~I lifetimes. The corresponding exponents $m$ 
are indicated in every panel. 
Second interesting feature in Figure~\ref{fig4} is seen along the line of constant 
$M_\ast$---there is a trend for models with greater ratio of rotational to gravitational energy
$\beta$ to have greater Class 0/I lifetimes. This effect is especially prominent for the Class~I objects
and is not unexpected. Cloud cores that have more energy stored in rotation
are expected to retain the envelope for a longer time due to an increased centrifugal force,
thus prolonging the embedded phase of star formation.

\begin{figure}
  \resizebox{\hsize}{!}{\includegraphics{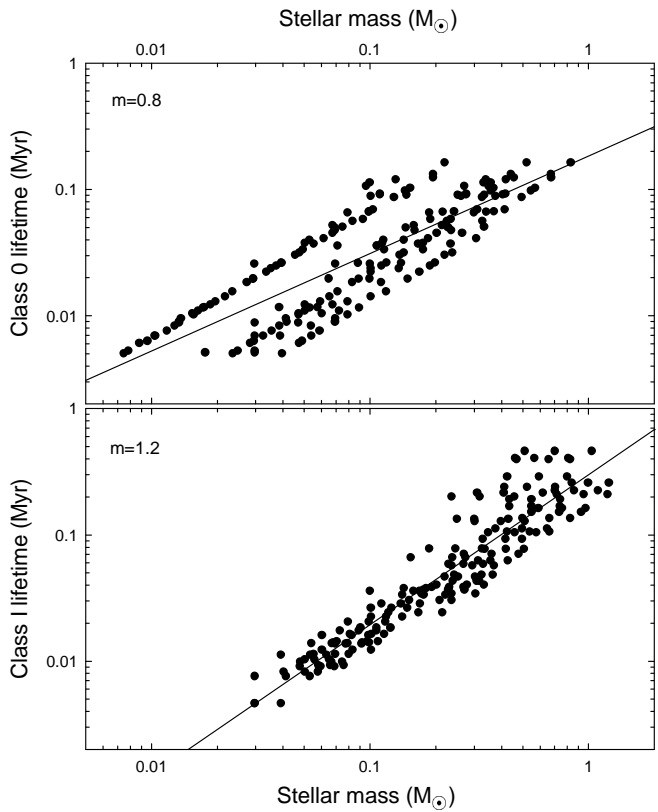}}
      \caption{Lifetimes of the Class~0 phase (top) and Class~I phase (bottom) versus stellar masses,
      the latter including those calculated at the beginning, in the midway, and at the 
      end of each evolution
      phase. The top/bottom panels of the figure are obtained by combining the top/bottom rows of Figure~\ref{fig4}.
      Solid lines present the least-squares best fits with exponents $m$.}
         \label{fig5}
\end{figure}

The correlations shown in each panel of Figure~\ref{fig4} may not be observationally 
meaningful when taken separately, because
they are derived for stars of specific age. As noted above, observations would not be biased towards
any specific instance of protostellar evolution but are likely to detect a whole spectrum of stellar
ages. To make our predictions more observationally significant, we plot in Figure~\ref{fig5} the Class~0
(top) and Class~I (bottom) lifetimes versus {\it all} stellar masses, namely, those at the beginning,
in the midway, and at the end of each evolution phase. In other words, in the top/bottom panel of Figure~\ref{fig5}
we combine the three top/bottom panels of Figure~\ref{fig4}. 
The least-squares best fits (solid lines) yield the following relations between 
Class~0 ($\tau_{\rm C0}$) and Class~I ($\tau_{\rm CI}$) 
lifetimes (in Myr), from one hand, and the corresponding stellar masses (in $M_\odot$), from the other hand
\begin{eqnarray}
\label{corr1}
\tau_{\rm C0} &=&  0.18 \, M_{\rm \ast,C0}^{0.8 \pm 0.05}, \label{correlation1}  \\ 
\label{corr2}
\tau_{\rm CI} &=&  0.30 \,M_{\rm \ast,CI}^{1.2 \pm 0.05}.
\label{correlation2}
\end{eqnarray}

It is evident that a somewhat steeper than linear correlation between $\tau_{\rm CI}$ and 
$M_{\rm \ast,CI}$ persists irrespective of the stellar age, i.e., independent of whether 
we measure stellar masses
in the beginning, in the midway, or at the end of the Class~I phase. This is a direct consequence
of the fact that a substantial fraction of the final stellar mass has already been accumulated
by the beginning of the Class~I phase and the subsequent increase in $M_{\rm \ast,CI}$ is not significant
(see Figure~\ref{fig3}). On the other hand, the Class~0 lifetimes demonstrate a noticeably weaker
correlation with stellar masses. This is because the latter vary
substantially during the Class~0 phase, which effectively spreads stellar masses along the horizontal
axis in Figure~\ref{fig5}, thus lowering the exponent in the least-squares fit.

This correlation between lifetimes and stellar masses suggests that the disagreement
between the observationally inferred Class~0 and Class~I lifetimes of YSOs may in part be due to variations
in the initial mass function or due to observational biases toward a particular mass band. 
To illustrate this idea, we calculate mean lifetimes for objects in our entire 
mass band  (0.008--0.85~$M_\odot$ and 0.03--1.24~$M_\odot$ for Class~0 and Class~I
objects, respectively) and for objects in the truncated mass band (0.2--0.85~$M_\odot$ and 0.2--1.24~$M_\odot$
for Class~0 and Class~I objects, respectively),  
with the latter case being confined to intermediate and upper-mass stars only. 
The resultant mean lifetimes of the Class~0 objects are $\langle \tau_{\rm C0}\rangle=0.044$~Myr 
and $\langle \tau_{\rm C0}\rangle_{\rm tr}=0.086$~Myr, respectively. 
In the case of the Class I objects, we obtain $\langle \tau_{\rm CI}\rangle=0.09$~Myr 
and $\langle \tau_{\rm CI}\rangle_{\rm tr}=0.15$~Myr.
It is evident that the neglect of the low-mass objects  causes almost a factor of 
two overestimate of the Class 0/I lifetimes. In spite of the fact that $\tau_{\rm C0}$
show a somewhat weaker correlation with $M_{\rm\ast,C0}$, the effect of truncation is even
greater than in the case of Class~I lifetimes due to systematically smaller stellar masses in the Class~0
phase.

\subsection{The effect of varying initial conditions and magnetic fields}
\label{initcond}

\begin{figure*}
 \centering
  \includegraphics[width=18cm]{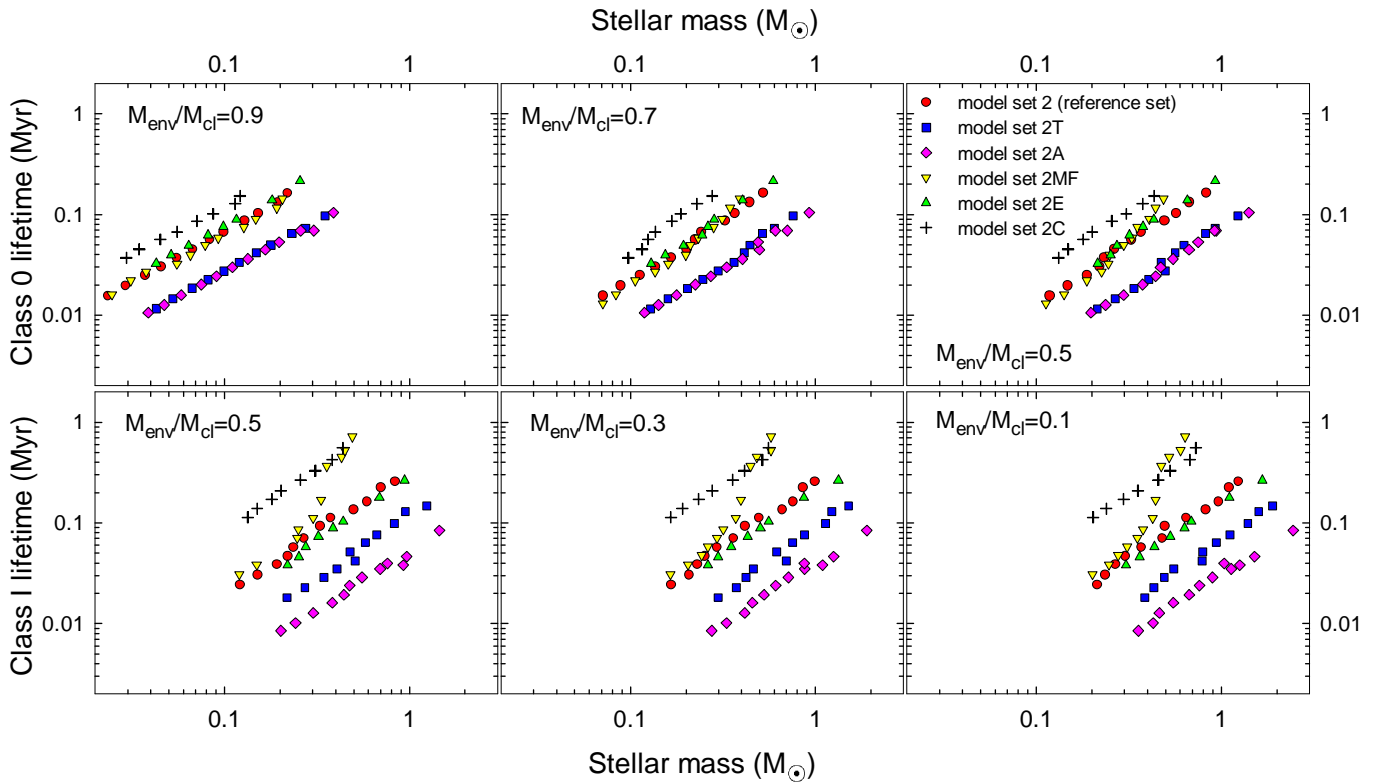}
      \caption{Lifetimes of the Class~0 phase (top row) and Class~I phase (bottom row) versus 
       stellar masses calculated at the beginning (right column), in the midway (middle column), and
       at the end (right column) of each evolution phase. 
       The data for six model sets with varying initial conditions (see Table~\ref{table1} 
       and the text) 
       are shown. The ratios $M_{\rm env}/M_{\rm cl}$ of the 
       envelope mass to the initial cloud core mass are indicated in every panel. }
         \label{fig6}
\end{figure*}

In the previous section, we have considered model cores that differ in the adopted 
values of rotational to gravitational energy $\beta$, initial mass $M_{\rm cl}$, 
and radius $r_{\rm out}$.
However, cloud cores may vary in other aspects such as initial gas temperatures, 
truncation radii, initial density enhancements, and strengths of magnetic fields. In addition,
the shape of the initial gas surface density profile may differ from the assumed scaling $\Sigma \propto
r^{-1}$.
It is computationally prohibitive to consider the effect of varying initial conditions on every model
core of Section~\ref{lifetimes}. Therefore, we have chosen model set~2 as the reference set 
to estimate the effect that the varying initial conditions may have on the lifetime
of the embedded phase.

Figure~\ref{fig6} shows our model lifetimes of the Class~0 phase (top panels) and
Class~I phase (bottom panels) versus stellar masses, the latter derived as described in the previous
section, for five model sets.  In particular, red
circles correspond to model set~2, blue squares correspond to model set~2T, which
is characterized by a higher initial gas temperature of the cloud cores $T=18$~K. 
Green triangles-up present model set 2E, which is 
characterized by extended cloud cores with $r_{\rm out}/r_0=12$ ($r_{\rm out}$ is increased). 
Pink diamonds show model set 2A,
in which cloud cores are characterized by the initial density enhancement amplitude $A=8$,
and yellow triangles-down correspond to model set 2MF with the flux-to-mass ratio $\alpha_{\rm m}=0.3$.
Finally, plus signs present model set~2C, which comprises models with spatially 
constant surface density and angular
velocity distributions. The latter are obtained using the parameters of the reference model set
and setting $\Sigma=\Sigma(r=r_{\rm out})$ and $\Omega=\Omega(r=r_{\rm out})$ 
in Equations~(\ref{dens}) and (\ref{omega}), 
respectively. In order words, surface densities and angular velocities in model set~2C
are set equal to those at the boundary  in the corresponding models of the reference set. 
The ratio $\beta=2.3\times 10^{-3}$ is the same for all models in this section in order to
exclude a mild dependence of the lifetimes on $\beta$ discussed in Section~\ref{lifetimes}.

It is evident that the varying initial conditions may have a profound effect on the
duration of the embedded phase. In particular, higher initial gas temperatures 
and density enhancement amplitudes reduce the resulting Class~0/I lifetimes by factors of 2.5--5.5
for objects with equal mass $M_\ast$.
This is not surprising, since both effects are expected to increase the mass accretion rate
from the envelope $\dot{M}_{\rm env}$, thus shortening Class~0/I lifetimes. 
In the standard model of inside-out collapse,  $\dot{M}_{\rm env}$ is 
proportional to the cube of the sound speed $c_{\rm s}^{3}$ 
and to the density enhancement amplitude $A$ \citep{Shu77}. In our case, the increase in gas 
temperature is $\triangle T=1.8$ 
and $\dot{M}_{\rm env}$ is expected to increase by a factor of $1.8^{3/2}=2.4$. Indeed, top panels 
in Figure~\ref{fig6} indicate that the corresponding Class~0 and Class~I lifetimes (blue squares) 
have decreased on average by factors of 2.5 and 3.2, respectively, as compared to the reference
set of models (red circles). A somewhat greater drop in $\tau_{\rm CI}$ than expected indicates
that the envelope empties in the late embedded phase faster than predicted by the standard model.
This could be due to the influence of the cloud core's finite outer boundary and an inward-propagating
rarefaction wave (cloud cores in the standard model are infinite in size). This effect
becomes even more pronounced in the case of a greater initial density enhancement $A=8$ 
(as opposed to $A=2$ in the reference
set). The corresponding Class~0 and Class~I lifetimes, shown by the pink diamonds 
in Figure~\ref{fig6}, have dropped by factors of 2.6 and 5.5, respectively. 

On the other hand, YSOs formed from cloud cores with greater truncation 
radii $r_{\rm out}/r_0=12$ (green triangles-up in Figure~\ref{fig6})  
do not seem to have significantly different lifetimes as compared to those of the reference 
model set with $r_{\rm out}/r_0=6$. 
This may seem counterintuitive. Indeed, more extended cores have a greater mass reservoir and
are supposed to deposit their material onto a star/disk system for a longer time. However, one has 
to keep in mind that they would also form more massive stars. As a result, the  
lifetimes simply shift toward the upper-right corner in every panel of Figure~\ref{fig6}.

YSOs formed from cloud cores with frozen-in magnetic fields ($\alpha_{\rm m}=0.3$) reveal a complicated
behavior. On one hand, their Class~0 lifetimes, as shown by yellow triangles-down 
in the top-left panel of Figure~\ref{fig6}, do not seem to be significantly affected. 
On the other hand, Class~I lifetimes of the intermediate- and upper mass stars increase 
considerably 
as compared to the reference set of model. The reason why these stars are more affected than their 
lower-mass counterparts
lies in the way that the frozen-in magnetic fields operate in cloud cores.  Let us consider the outermost
layer of gas. As it moves toward the center, magnetic tension increases
and decelerates the collapse, prolonging the embedded phase. Magnetic field lines are attached
at infinity to the natal giant molecular cloud, which evolves on a much slower time scale than the
collapsing cloud core. This effectively means that the field lines are motionless at infinity. 
It now follows that magnetic tension would be stronger for larger
(and hence more massive) cores, as it takes a larger distance for the outermost layer to
travel toward the disk surface. If we recall that more massive cores 
produce more massive stars, a steeper dependence of $\tau_{\rm CI}$ on $M_{\rm\ast,CI}$ 
becomes evident. We note that this effect may be either magnified 
if higher values of $\alpha_{\rm m}$ are present in cloud cores or  moderated for lower values of 
$\alpha_{\rm m}$. In addition, ambipolar diffusion and magnetic breaking, 
not included in the present numerical simulations, are expected to decrease the effect.

Finally, the initial distribution of gas surface density and angular velocity in cloud cores 
may also affect the resulting lifetimes. For the case of cloud cores with initially constant 
$\Sigma$ and $\Omega$ (plus signs in Figure~\ref{fig6}), the resulting lifetimes 
are greater than those in the reference model set, the latter having both $\Sigma$ and 
$\Omega$ inverse proportional to radius 
at large radii. The increase is particularly large in the case of Class~I lifetimes
and is caused by a steep radial gas surface density distribution
that develops in the late phases of cloud core collapse. Although cloud cores in model set~2C start
with constant $\Sigma$, they quickly evolve into a configuration that is characterized by a 
gas surface density profile that declines with radius steeper than $r^{-1}$ at large radii.
This acts to lower the mass infall rate from the envelope $\dot{M}=-2 \pi r v_{\rm r} \Sigma$ 
as compared to that in the reference model, the latter being characterized by 
$\Sigma\propto r^{-1}$  at $r>>r_0$. The effect is particularly strong in the late evolution
when gas layers initially located at large radial distances begin to fall in onto the disk.

How do varying initial conditions affect the correlation between Class~0/I 
lifetimes and masses of the central object discussed in the previous section? 
To quantitatively answer this question would require to run a prohibitively large
number of models for every model cloud core with all possible initial conditions. However,
based on the behaviour of model cores considered in this section, we can assume that 
the correlations given by Equations~(\ref{corr1}) and (\ref{corr2}) are unlikely to break 
completely. 
Indeed, in most model sets with varying initial conditions (apart from that with non-zero
frozen-in magnetic fields) the slope of the correlation line is similar to that of model set~2.
It is thus the spread in the lifetimes that can affect somewhat the
resulting correlation. Just to estimated the effect, we have combined 
Class~0 and Class~I data from {\it all} models into two separate
$\tau$--$M_\ast$ plots for each phase and found that the resulting least-squares exponents
in Equations~(\ref{corr1}) and (\ref{corr2}) decrease by as much as 0.3.

\begin{figure}
  \resizebox{\hsize}{!}{\includegraphics{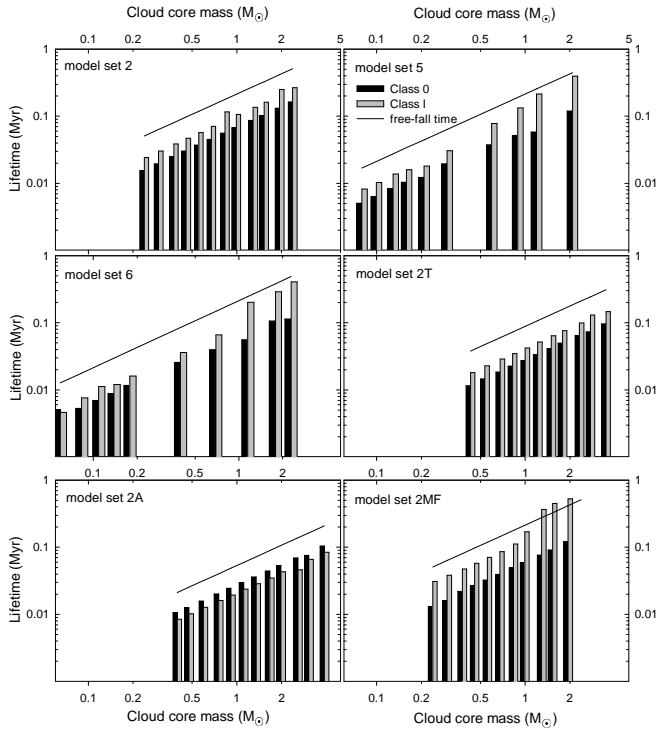}}
      \caption{Class~0 (black bars) and Class~I (grey bars) lifetimes versus initial 
      cloud core masses for six model sets as specified in each panel. Solid lines show the 
      free-fall time $\tau_{\rm ff,out}$ of the outermost gas layer as derived from Equation~(\ref{tauff}).}
         \label{fig7}
\end{figure}

\subsection{Class~I lifetime versus Class~0 lifetime}
\label{lifetimeratios}
It is often useful to know how Class~0 and Class~I lifetimes are related to each other 
for a {\rm particular} YSO. It is difficult to extract such information directly 
from Figures~\ref{fig4}--\ref{fig6} and  Equations~(\ref{corr1}) and (\ref{corr2}) are not 
useful because they provide data averaged over all model cores. 
In order to analyse the relative lifetimes, we have to choose some characteristic of a YSO 
in order to present $\tau_{\rm C0}$ and 
$\tau_{\rm CI}$ as a function of this characteristic.
We use the initial cloud core masses with a practical purpose
to verify if the lifetimes indeed scale linearly with $M_{\rm cl}$, as was suggested on theoretical
grounds at the beginning of Section~\ref{justify}.

Figure~\ref{fig7} presents the Class~0 (black bars) and Class~I (grey bars) 
lifetimes as a function of the initial cloud core mass $M_{\rm cl}$ for model cores from 
six model sets as specified in each panel. Every pair of bars represents an individual YSO 
formed from a cloud core with mass $M_{\rm cl}$.
It is evident that $\tau_{\rm CI}$ are systematically greater than 
$\tau_{\rm C0}$, except for model set~2A for which
$\tau_{\rm CI}<\tau_{\rm C0}$. We note that the latter case is characterized by a rather 
large initial density enhancement $A=8$ and need specific  conditions (such as strong shock 
waves) to be realized. It is also seen that the ratio $\tau_{\rm CI}/\tau_{\rm C0}$
varies little with $M_{\rm cl}$, with a notable exception for models with 
magnetic fields (model set 2MF) and for models with large rotation rates 
(model set 5 and 6). We have already discussed above the case with magnetic
fields. As to the models with greater values of the rotational to gravitational energy 
$\beta$, they have more centrifugal support against gravity, which effectively prolongs the Class~I
phase in the upper-mass (and spatially extended) cloud cores. Excluding model set~2A and 
the upper mass models in model sets~5, 6, and 2MF, the mean 
ratio of the Class~I to the Class~0 lifetime is $\tau_{\rm CI}/\tau_{\rm C0}\approx$1.5--2.0.
For the aforementioned upper-mass models, the corresponding ratio may be as high as 3.0--4.0.

The solid lines in Figure~\ref{fig7} show the free-fall time $\tau_{\rm ff,out}$ of the cloud's 
outermost layer as a function of $M_{\rm cl}$. The 
free-fall time is linearly proportional to $M_{\rm cl}$ (see equation~(\ref{tauff}) 
in the Appendix). By comparing lifetimes with the free-fall time,
one can notice that both the Class~0 and Class~I lifetimes in model sets~2, 2T, and 2A are 
also linearly proportional 
to $M_{\rm cl}$, as was anticipated from simple theoretical grounds in Section~\ref{justify}. 
However, the Class~I lifetimes of the upper-mass model cores in model 
sets 5, 6, and 2MF show a significant deviation from the linear scaling typical for 
the low- and intermediate-mass models. This effect was not envisioned from simple 
theoretical considerations.

\section{Envelope depletion rates}
\label{depletion}
In semi-analytic models of cloud core collapse and star/disk formation, 
it is often needed to know the rate at which 
the envelope material is accreted onto the star/disk system. 
We determine this rate by calculating the envelope depletion rate $\dot{M}_{\rm env}= 
- \left( M_{\rm env}(t+\triangle t) - M_{\rm env}(t) \right)/\triangle t$, with the time step 
$\triangle t$ set to $10^3$~yr.  In the earliest stages of evolution before the disk formation, 
$\dot{M}_{\rm env}$ would represent the mass accretion rate onto the star, but when the disk forms 
$\dot{M}_{\rm env}$ would represent the mass accretion rate onto the disk.

\begin{figure*}
 \centering
  \includegraphics[width=18cm]{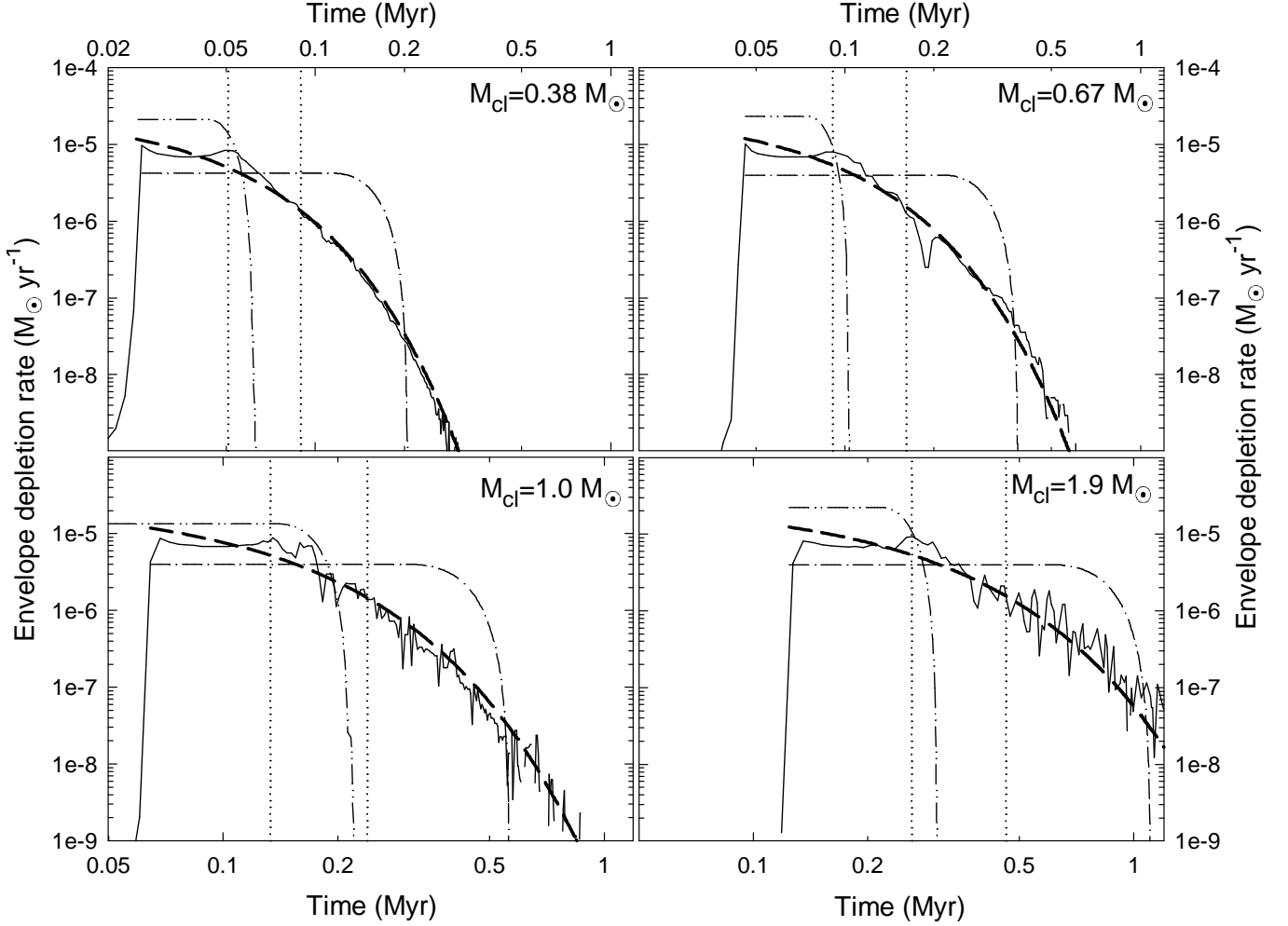}
      \caption{Envelope depletion rates $\dot{M}_{\rm env}$ 
      (or, equivalently, mass accretion rates onto a star/disk system)
      versus time elapsed since the beginning of numerical simulations for
      four representative models from model set~2. The corresponding cloud core masses
      are shown in each panel. The solid lines show the model data, while the thick 
      dashed lines are 
      derived using Equation~(\ref{mybatc}). The dashed-dotted and dash-dot-dotted lines
      show $\dot{M}_{\rm env}$ as derived using the RMA function~(\ref{rma}) with $\tau_{\rm ff}=\tau_{\rm
      ff,out}$ and $\tau_{\rm ff}=\tau_{\rm ff,2/3}$, respectively (see the text for details). The vertical
      dotted lines mark the onset of the Class~I (left) and Class~II (right) phases. }
         \label{fig8}
\end{figure*}

The solid lines in Figure~\ref{fig8} show the envelope depletion rate versus time elapsed since 
the beginning of numerical simulations for the same four representative models as 
in Figure~\ref{fig3}. The vertical
dotted lines mark the onset of the Class I (left line) and Class II (right line) phases.
The time evolution of $\dot{M}_{\rm env}$ can be split into two distinct modes: a shorter period 
of near-constant depletion rate and a longer period of gradual (and terminal) decline 
of $\dot{M}_{\rm env}$. It is seen that the 
boundary between these two modes lies near the vertical dotted line that separates 
the Class~0 and Class I phases of star formation in every model. A similar behavior of $\dot{M}_{\rm
env}$ was reported by \citet{VB05a} for the case of the {\it spherically} symmetric 
collapse of truncated cloud cores. 

In the first mode, the accretion process is
conceptually similar to that described by the single isothermal sphere solution \citep{Shu77}, wherein
the material is falling freely onto the star/disk system and the accretion rate is constant 
and proportional to $c_{\rm s}^3/G$. 
The second mode ensues when a rarefaction
wave propagating inward from the core's outer boundary hits the disk surface. From this moment on, 
the accretion process is no more self-similar, the radial gas density profile starts to steepen 
and the ongoing depletion of gas in the envelope causes a gradual (and terminal) 
decline of $\dot{M}_{\rm env}$. 

We note that $\dot{M}_{\rm env}$ may exhibit some short-time 
fluctuations in the second mode due to temporary disk expansions and contractions 
caused by disk radial pulsations\footnote{The animation of this process can be viewed
at www.astro.uwo.ca/$\sim$vorobyov (animations: burst mode of accretion)}. When expanding, the disk
delivers part of its material to the envelope, thus temporarily decreasing the envelope depletion 
rate $\dot{M}_{\rm env}$. The fluctuation amplitudes grow along the line of increasing cloud core 
masses (and, consequently, increasing disk masses), indicating that gravitational instability 
lies behind these fluctuations.
These fluctuations, though having a conceptually different nature, are similar in behavior to 
the spasmodic accretion phenomenon reported by \citet{Tassis05}.

We now want to find a simple and convenient way to approximate our model envelope depletion
rates. We consider two different functions: the first one proposed by \citet{Bontemps96} to 
account for the observed outflow energetics in the embedded phase (hereafter, 
the BATC function) and the second one employed by \citet{Rice09} to mimic the
mass infall rates onto the star/disk system in their one-dimensional global simulations of cloud core
collapse (hereafter, the RMA function)
\begin{equation}
\dot{M}_{\rm BATC}(t) = {M_{\rm cl} \over \tau_{\rm c}} \times e^{-t/\tau_{\rm c}},
\label{batc}
\end{equation}
\begin{equation}
\dot{M}_{\rm RMA}(t)= \left\{ \begin{array}{ll}
{M_{\rm cl} \over \tau_{\rm ff} }  & \,\,\,  0\le t \le \tau_{\rm ff}  \\
{1\over 2}{M_{\rm cl}\over \tau_{\rm ff}} \left[ 1 + \cos{\pi(t-\tau_{\rm ff}) \over \tau_{\rm ff}}\right]& \,\,\,  \tau_{\rm ff} \le t \le 2 \tau_{\rm ff} \label{rma} \\
0 & \,\,\, 2\tau_{\rm ff}\le t.
\end{array}   \right.
\label{rma}
\end{equation}
Here, $\tau_{\rm c}$ is some characteristic time, the value of which was constrained by 
\citet{Bontemps96} using the inferred duration
of the embedded phase, and $\tau_{\rm ff}$ is the free-fall time.

First, we check the utility of the BATC function.
When one undertakes modeling of the gravitational collapse of a cloud core, the
value of $\tau_{\rm c}$ is not known {\it a priori}.
Therefore, it is desirable to link $\tau_{\rm c}$ with the free-fall time
$\tau_{\rm ff}=(3\pi/32 G \rho)^{1/2}$.
We have experimented with various values of $\tau_{\rm ff}$, depending on the actual value 
of $\rho$, and 
recommend the following relation $\tau_{\rm c}=\tau_{\rm ff,out}/3$, 
where $\tau_{\rm ff,out}$ is the free-fall time of the outermost layer 
derived in the Appendix. The time $t$ in Eqs.~(\ref{batc}) and (\ref{rma}) 
is counted from the formation of the protostar (the accretion rate is negligible 
at the earlier time).
The resulting mass accretion rate $\dot{M}_{\rm BATC}$ is shown in Figure~\ref{fig8}
by the thick dashed lines. It is evident that the BATC function provides an adequate fit to our
model accretion rates (solid lines). This result is robust and holds for 
models with different 
values of $\beta$ and with different initial conditions.

We now proceed with considering the utility of the RMA function. Since our cloud
cores are initially non-uniform, we have considered several values of $\tau_{\rm ff}$,
depending on the actual value of $\rho$. The dash-dotted and dash-dot-dotted
lines in Figure~\ref{fig8} show
the resulting mass accretion rate $\dot{M}_{\rm RMA}$ for $\tau_{\rm ff}=\tau_{\rm ff,out}$
and $\tau_{\rm ff}=\tau_{\rm ff,2/3}=\left( 3 \pi /(32 G \rho_{\rm 2/3}) \right)^{1/2}$, respectively, where the latter value is the free-fall time
of a layer located at $r=2r_{\rm out}/3$ (i.e. at a radial distance equal to 2/3 that of the cloud
core outer radius) and $\rho_{\rm 2/3}$ is the corresponding gas volume density. 
It is obvious that, regardless of the value of $\tau_{\rm ff}$, 
the RMA function fails to provide an acceptable fit to our model accretion rates (solid lines).
To summarize, we suggest using the following equation as the mass accretion rate onto a star/disk
system
\begin{equation}
\dot{M}_{\rm env}(t) = { 3 M_{\rm cl} \over \tau_{\rm ff,out}} \times e^{- 3 t/\tau_{\rm ff,out}},
\label{mybatc}
\end{equation}
where time $t$ should be counted from the formation of a central star and $\tau_{\rm ff,out}$ (see
Equation~[\ref{tauff}])
is the free-fall time of the outermost gas layer of a cloud core with mass $M_{\rm cl}$.

\section{Conclusions}
\label{summary}
Using numerical hydrodynamics simulations, we have computed 
the gravitational collapse of a large set of cloud cores. We start from a gravitationally 
unstable, pre-stellar phase and terminate the simulations in the Class~II phase when
most of the cloud core has been accreted by the forming star/disk system.
We have considered cloud cores with various initial masses ($M_{\rm cl}$=0.06--3.9~$M_\odot$), 
ratios of the rotational to the gravitational energy ($\beta$=(0.2--2.2)$\times10^{-2}$), 
initial gas temperatures ($T_0$=10--18~K), truncation radii ($r_{\rm out}/r_0$=6--12), 
strengths of frozen-in magnetic fields ($\alpha_{\rm m}$=0--0.3), initial gas 
density enhancements ($A$=2--8), and initial radial profiles of the gas surface density and angular
velocity ($\Sigma,\Omega \propto r^{-1}$ and $\Sigma, \Omega =const$).

We employ a sophisticated method for distinguishing between 
the infalling envelope  and the forming disk in our numerical simulations 
and determine the duration of the
embedded phase of star formation, adopting a classification scheme based on the remaining 
mass in the envelope \citep{Andre93}. We also calculate
the envelope depletion rate $\dot{M}_{\rm env}$ (or, equivalently, the rate of mass accretion 
onto the star/disk system) and test the utility of two empirical functions for 
$\dot{M}_{\rm env}$ provided by \citet{Bontemps96} (BATC function)  and \citet{Rice09} (RMA function).
We find the following.
\begin{enumerate}
\item 
Class~0 ($\tau_{\rm C0}$) and Class~I ($\tau_{\rm CI}$) lifetimes correlate with the 
corresponding stellar mass $M_{\rm \ast,C0}$ and $M_{\rm \ast,CI}$, 
however the scaling is more complex than can 
be expected from simple theoretical grounds based on the linear correlation between the 
free-fall time and the cloud core mass.
In particular, the form of the correlation depends on the cloud core properties
such as the initial ratio of rotational to gravitational energy, initial gas temperature, 
density enhancement amplitude, and the initial radial distribution of gas surface density
and angular velocity.  In addition, frozen-in magnetic fields with constant 
flux-to-mass ratio can substantially influence the Class~I lifetimes. The correlation is further
complicated due to the fact that the stellar mass varies considerably during the Class~0 phase. 
This makes the $\tau_{\rm C0}$--$M_{\rm \ast,C0}$ correlation dependent on the stellar age as well.
\item 
When cloud cores with varying rotation rates, masses and sizes (but otherwise identical) are 
considered,  Class~0  and Class~I lifetimes  have sub- and super-linear
correlations with the corresponding stellar masses, respectively. These correlations extend from
sub-stellar masses ($\sim 0.01~M_\odot$) to at least solar masses ($\sim 1.5~M_\odot$) 
and obey the following
scaling laws $\tau_{\rm C0}=0.18~M_{\rm \ast,C0}^{0.8\pm0.05}$ and 
$\tau_{\rm CI}=0.3~M_{\rm \ast,CI}^{1.2 \pm 0.05}$
for Class~0/I stars of {\it all possible ages}.
This makes the observational determination of the mean lifetimes sensitive to the form 
of the initial mass function and/or to instrumental biases toward a particular mass band. 
For instance, our modeling predicts a mean Class~0 lifetime of
 $\langle \tau_{\rm C0}\rangle =0.044 $~Myr for
stars in the 0.008--0.85~$M_\odot$ mass range and $\langle \tau_{\rm C0,tr}\rangle =0.086 $~Myr
for stars in the  0.02--0.85~$M_\odot$ range. 
In the case of the Class I objects, we obtain $\langle \tau_{\rm CI}\rangle=0.09$~Myr 
and $\langle \tau_{\rm CI}\rangle_{\rm tr}=0.15$~Myr for stars in the 0.03--1.24~$M_\odot$ 
and 0.2--1.24~$M_\odot$ mass ranges, respectively.
It is evident that the neglect of objects at the lower
mass end results in almost a factor of 2 overestimate of the mean lifetimes.

\item An increase in the initial cloud core temperature and density 
enhancement amplitude tend to lower the Class~0 and Class~I 
lifetimes, whereas cloud cores with initially constant gas surface density and angular
velocity distributions (as compared to those with $\Sigma,\Omega \propto r^{-1}$)
have longer Class~0 and especially Class~I lifetimes.
In addition, frozen-in magnetic fields may increase the Class~I lifetimes, 
particularly for the upper-mass cloud cores, thus steepening the corresponding 
$\tau_{\rm CI}$--$M_{\rm \ast,CI}$
relation. The net effect of varying initial condition is to
{\it weaken} the aforementioned sub-linear (Class~0) and super-linear (Class~I) 
correlations between the lifetimes and stellar masses 
by decreasing the corresponding exponents by as much
as 0.3. However, more accurate modeling with ambipolar diffusion and magnetic braking 
taken into account is needed to accurately assess the magnitude of this effect.
The outer truncation radius of a cloud core has little effect on the resulting lifetimes.

\item Most cloud cores give birth to YSOs whose Class~I lifetimes are longer than those
of the Class~0 phase by roughly a factor of 1.5--2.0.
A notable exception are YSOs formed from cloud cores a with large initial 
density enhancement. In the latter case, the duration of the
Class~I phase may actually be shorter than that of the Class~0. 
In addition, the upper-mass models $M_{\rm cl}>1~M_\odot$ with frozen-in magnetic fields 
and high cloud core rotation rates $\beta>10^{-2}$ may have the $\tau_{\rm CI}/\tau_{\rm C0}$ 
ratios as large as 3.0--4.0.

\item The time evolution of $\dot{M}_{\rm env}$ reveals two 
distinct modes:  a shorter period of near-constant depletion rate and a longer period 
of gradual (and terminal) decline of $\dot{M}_{\rm env}$. 
The boundary between these two modes lies near the end of the Class~0 phase and the beginning
of Class I phase. In the later mode, $\dot{M}_{\rm env}$ may show short-term fluctuations due to
episodic disk expansions and contractions. The BATC function may provide an adequate fit to our 
model envelope depletion rates if the characteristic time of decline of $\dot{M}_{\rm env}$
is chosen properly. The RMA function fails to provide an acceptable fit to
our model data, irrespective of the free parameters.

\end{enumerate}

We emphasize that our lifetimes have been derived based on the AWTB classification scheme \citep{Andre93}
and may change by a factor of unity if other schemes  are used. For instance, if we re-define the boundary
between the Class I and Class~II phases in the AWTB scheme and assume that the Class~II phase begins
when the envelope mass drops below 5\% of the initial cloud core mass (in contrast to 10\% adopted in
our paper), the resulting Class~I lifetimes in model set~2 increase by a factor of 1.5.
However, the derived trends and correlations between the lifetime of the embedded phase, 
from one hand, and the stellar masses, from the other hand,
are expect to stay valid irrespective of the classification scheme used.

\acknowledgments
   The author is thankful to the anonymous referee for suggestions and comments that helped to improve
   the final manuscript. The author gratefully acknowledges present support 
   from an ACEnet Fellowship and Prof. Rob Thacker. Numerical simulations were done 
   on the Atlantic Computational Excellence Network (ACEnet) and at the Center of Collective Supercomputer
   Resources, Taganrog Technological Institute at South Federal University.

\appendix

\section{Relation between the cloud core mass and the free-fall time}
\label{relation} 
If the initial gas volume density distribution is non-uniform, 
the usual definition of $\tau_{\rm ff}$ needs to be further clarified as to
what value of $\rho$ should be used. In our case, it is convenient to set $\rho=\rho_{\rm out}=
\rho(r=r_{\rm out})$,
i.e. we use the gas volume density at the outer cloud core boundary. Then, the resulting
free-fall time of the outermost gas layer becomes 
$\tau_{\rm ff,out}=\sqrt{3\pi/(32G\rho_{\rm out})}$. 

We now need to relate $\tau_{\rm ff, out}$ with  the initial cloud core mass $M_{\rm cl}$.
Neglecting a small plateau in the gas surface density profile at $r\la r_0$, we can write (see Section~\ref{init})
\begin{equation}
\Sigma={A^{1/2} c_{\rm s}^2 \over \pi G r}.
\end{equation}
The mass of a cloud core with this gas surface density distribution is 
\begin{equation}
M_{\rm cl}= {2 A^{1/2} c_{\rm s}^2 r_{\rm out} \over G}= {2 A c_{\rm s}^4 \over \pi G \Sigma_{\rm out}},
\label{mass}
\end{equation}
where $\Sigma_{\rm out}$ is the gas surface density at the cloud core's outer boundary.
Assuming a local vertical hydrostatic equilibrium with the scale height $Z=c_{\rm s}^2/(\pi G \Sigma)$,
approximating the gas volume density as $\rho = \Sigma/(2 Z)$, and substituting the resulting 
value of $\rho_{\rm out}=\Sigma_{\rm out}^2 \pi G/(2 c_{\rm s}^2)$ in the expression for $\tau_{\rm ff,out}$, we finally obtain
\begin{equation}
\tau_{\rm ff,out}[\mathrm{Myr}]={3.8 \over A} \left( {\mu \over  T_0} \right)^{3/2} \left( {M_{\rm cl} \over M_\odot} \right),
\label{tauff}
\end{equation}
where $T_0$ is the initial gas temperature, $\mu$ is the mean molecular weight, and $A$
is the amplitude of initial density enhancement.


\begin{thebibliography}{ }

\bibitem[\protect\citeauthoryear{Andr\'e et al.}{1993}]{Andre93}
Andr\'e, P., Ward-Thompson, D., \& Barsony, M. 1993, ApJ, 406, 122

\bibitem[\protect\citeauthoryear{Andr\'e \& Montmerle}{1994}]{Andre94}
Andr\'e, P., \& Montmerle, T. 1994, ApJ, 420, 837

\bibitem[\protect\citeauthoryear{Bacmann et al.}{2000}]{Bacmann00}
Bacmann, A., Andr\'e, P., Puget, J.-L., Abergel, A., Bontemps, S., 
Ward-Thompson, D., 2000, A\&A, 361, 555

\bibitem[\protect\citeauthoryear{Basu}{1997}]{Basu}
Basu, S. 1997, ApJ, 485, 240  

\bibitem[\protect\citeauthoryear{Boley et al.}{2009}]{Boley09}
Boley, A. C., Hayfield, T., Mayer, L., \& Durisen, R. H. 2009, astro-ph/09094543

\bibitem[\protect\citeauthoryear{Bontemps et al.}{1996}]{Bontemps96}
Bontemps, S., Andr\'e, P., Terebey, S., Cabrit, S. 1996, A\&A, 311, 858

\bibitem[\protect\citeauthoryear{Caselli et al.}{2002}]{Caselli}
Caselli, P., Benson, P. J., Myers, P. C., \& Tafalla, M. 2002, 572, 238

\bibitem[\protect\citeauthoryear{Dapp \& Basu}{2009}]{Wolf09}
Dapp, W. B., Basu, S. 2009, MNRAS, 395, 1092.

\bibitem[\protect\citeauthoryear{Dunham et al.}{2010}]{Dunham10}
Dunham, M. M., Evans II, N. J., Terebey, S., Dullemond, P., Young, C. H. 2010, to appear in ApJ

\bibitem[\protect\citeauthoryear{Enoch et al.}{2006}]{Enoch06}
Enoch, M. L., Eet al. 2006, ApJ, 638, 293

\bibitem[\protect\citeauthoryear{Enoch et al.}{2009}]{Enoch09}
Enoch M. L., Evans II, N. J., Sargent, A. I., \& Glenn, J. 2009, ApJ, 692, 973

\bibitem[\protect\citeauthoryear{Evans et al.}{2009}]{Evans09}
Evans II, N. J., Dunham, M. M., Jorgensen, J. K., Enoch, M. L., Mer\'in, B., van Dishoeck,
E. F., et al. ApJSS, 181, 321

\bibitem[\protect\citeauthoryear{Froebrich et al.}{2006}]{Froebrich06}
Froebrich, D., Schmeja, S., Smith, M. D., \& Klessen, R. S. 2006, MNRAS, 368, 435

\bibitem[\protect\citeauthoryear{Greene et al.}{1994}]{Greene94}
Greene, T. P., Wilking, B. A., Andr\'e, P., Young, E. T., \& Lada, C. J. 
1994, ApJ, 434, 614

\bibitem[\protect\citeauthoryear{Hatchell et al.}{2007}]{Hatchell07}
Hatchell, J., Fuller, G. A., Richer, J. S., Harries, T. J., \& Ladd E. F. 
2007, A\&A, 468, 1009

\bibitem[\protect\citeauthoryear{Kenyon \& Hartmann}{1995}]{Kenyon95}
Kenyon, S. J., \& Hartmann, L. 1995, ApJS, 101, 117

\bibitem[\protect\citeauthoryear{Kratter et al.}{2009}]{Kratter09}
Kratter, K. M., Matzner, C. D., Krumholz, M. R., \& Klein, R. I., astro-ph:0907.3476


\bibitem[\protect\citeauthoryear{Lada}{1987}]{Lada87}
Lada, Ch. J. 1987, in IAU Symp. 115, Star Forming Regions, ed. M. Peimbert
\& J. Jugaku (Dordrecht: Reidel), 1

\bibitem[\protect\citeauthoryear{Larson}{2003}]{Larson}
Larson, R. B., 2003, Rep. Prog. Phys., 66, 1651

\bibitem[\protect\citeauthoryear{Lin \& Pringle}{1990}]{LP90}
Lin, D.N.C., Pringle, J.E. 1990, ApJ 358, 515


\bibitem[\protect\citeauthoryear{Masunaga \& Inutsuka}{2000}]{Masunaga00}
Masunaga, H., \& Inutsuka, S. 2000, ApJ, 531, 350


\bibitem[\protect\citeauthoryear{Myers \& Ladd}{1993}]{Myers93}
Myers, P. C., \& Ladd, E. F. 1993, ApJ, 413, L47

\bibitem[\protect\citeauthoryear{Rice et al.}{2009}]{Rice09}
Rice, W. K. M., Mayo, J. H., \& Armitage, P. J. 2009, astro-ph:0911.1202

\bibitem[\protect\citeauthoryear{Shu}{1977}]{Shu77}
Shu, F. H. 1977, ApJ, 214, 488

\bibitem[\protect\citeauthoryear{Tassis \& Mouschovias}{2005}]{Tassis05}
Tassis, K., \& Mouschovias, T. C. 2005, ApJ, 618, 783

\bibitem[\protect\citeauthoryear{Visser et al.}{2002}]{Visser02}
Visser, A. E., Richer, J. S., Chandler, C. J. 2002, AJ, 124, 2756

\bibitem[\protect\citeauthoryear{Vorobyov \& Basu}{2005}]{VB05a}
Vorobyov, E. I., \& Basu, S. 2005, MNRAS, 360, 675


\bibitem[\protect\citeauthoryear{Vorobyov \& Basu}{2006}]{VB06}
Vorobyov, E. I., \&  Basu, S. 2006, ApJ, 650, 956

\bibitem[\protect\citeauthoryear{Vorobyov}{2009a}]{Vor09}
Vorobyov, E. I., 2009a, ApJ, 692, 1609 

\bibitem[\protect\citeauthoryear{Vorobyov}{2009b}]{Vor09b}
Vorobyov, E. I. 2009b, ApJ, 704, 715

\bibitem[\protect\citeauthoryear{Vorobyov}{2009c}]{Vor09c}
Vorobyov, E. I. 2009c, New Astronomy, 15, 24

\bibitem[\protect\citeauthoryear{Vorobyov \& Basu}{2009a}]{VB09a}
Vorobyov, E. I., \& Basu, S. 2009a, MNRAS, 393, 822

\bibitem[\protect\citeauthoryear{Vorobyov \& Basu}{2009b}]{VB09b}
Vorobyov, E. I., \& Basu, S. 2009b, ApJ, 703, 922



\bibitem[\protect\citeauthoryear{Wilking et al.}{1989}]{Wilking89}
Wilking, B. A., Lada, C. J., \& Young, E. T. 1989, ApJ, 340, 823


\end{thebibliography}
\end{document}